\documentclass[runningheads]{llncs}
\usepackage{graphicx}
\usepackage{url}
\usepackage{amsmath}
\usepackage{comment} 
\usepackage{listings}
\usepackage{xcolor}
\usepackage{soul}
\usepackage{float}
\usepackage{courier}
\usepackage{makecell}
\usepackage{tabularx}

\usepackage{hyperref}
\usepackage{seqsplit}
\usepackage[english]{babel}
\usepackage{hyphenat}
\usepackage{listingsutf8}
\usepackage{enumitem}
\usepackage[misc,geometry]{ifsym} 
\usepackage[normalem]{ulem}
\usepackage{caption}
\DeclareCaptionFont{ninept}{\fontsize{9pt}{11pt}\selectfont #1}
\setlist[itemize]{noitemsep, nolistsep}
\setlist[enumerate]{noitemsep, nolistsep}

\usepackage{upquote}

\lstdefinelanguage{SPARQL2}{
  language     = SPARQL,
  morekeywords = {IF, BIND, SERVICE, AS, CONCAT, STRAFTER, IRI, \{,\}},
}

\begin{document}

\title{VoIDext:\thinspace Vocabulary and Patterns for Enhancing Interoperable Datasets with Virtual Links}
\titlerunning{VoIDext - an RDF schema vocabulary}
%
\author{Tarcisio M. de Farias\inst{1,2}(\Letter) \and
Kurt Stockinger\inst{4} \and
Christophe~Dessimoz\inst{1,2,3}
}
\authorrunning{T. Mendes de Farias, K. Stockinger and C. Dessimoz}
%
\institute{SIB Swiss Institute of Bioinformatics, Lausanne, Switzerland \\ 
\and 
University of Lausanne, Switzerland \\ 
 \and  
University College London, UK  \\ 
\and
Zurich University of Applied Sciences, Winterthur, Switzerland
\email{\{tarcisio.mendesdefarias\}@unil.ch}
}
\maketitle              
\begin{abstract}

 Semantic heterogeneity remains a problem when interoperating with data from  sources of different scopes and knowledge domains. Causes for this challenge are context-specific requirements (i.e. no ``one model fits all"), different data modelling decisions, domain-specific purposes, and technical constraints. Moreover,  even if the problem of semantic heterogeneity among different RDF publishers and knowledge domains is solved, querying and accessing the data of distributed RDF datasets on the Web is not straightforward. This is because of the complex and fastidious process needed to understand how these datasets can be related or linked, and consequently, queried. To address this issue, we propose to extend the existing Vocabulary of Interlinked Datasets (VoID) by introducing new terms such as the {\it Virtual Link Set} concept and data model patterns. A virtual link is a connection between resources such as literals and IRIs (Internationalized Resource Identifier) with some commonality where each of these resources is from a different RDF dataset. The links are required in order to understand how to semantically relate datasets. In addition, we describe several benefits of using virtual links to improve interoperability between heterogenous and independent datasets. Finally, we exemplify and apply our approach to multiple world-wide used RDF datasets.

\end{abstract}
\section{Introduction}
To achieve semantic and data interoperability, several data standards, ontologies, thesauri, controlled vocabularies, and taxonomies have been developed and adopted both by academia and industry. For example, the Industry Foundation Classes \cite{ifc} is an ISO standard to exchange data among Building Information Modelling software tools \cite{farias_2018}. In life sciences, we can mention the Gene Ontology (GO) among many other ontologies listed in repositories such as BioPortal \cite{whetzel_2011}. Yet, semantic heterogeneity remains a problem when interoperating with data from various sources which represent the same or related information in different ways \cite{farias_2015a}. This is mainly due to the lack or difficulty of a common consensus, different modelling decisions, domain scope and purpose, and constraints (e.g. storage, query performance, legacy and new systems).

Semantic reconciliation---i.e. the process of identifying and resolving semantic conflicts \cite{Siegel1991-ui}, for example, by matching concepts from heterogeneous data sources \cite{Gal2005-ko}---is recognized as a key process to address the semantic heterogeneity problem. To support this process, ontology matching approaches \cite{otero_2015} have been proposed such as YAM++ \cite{ngo_2016}. Although semantic reconciliation enhances semantic interoperability, it is often not fully applicable or practical when considering distributed and independent RDF (Resource Description Framework) datasets of different domain scopes, knowledge domains, and autonomous publishers. In addition, even if the semantic reconciliation process among different RDF publishers and knowledge domains is complete and possible, querying and accessing the data of multiple distributed RDF datasets on the Web is not straightforward. This is because of the complex, time-consuming and fastidious process of having to understand how the data are structured and how these datasets can be related or linked, and consequently, queried.

To enhance interoperability and to facilitate the understanding of how multiple datasets can be related and queried, we propose to {\it extend and adapt the existing Vocabulary of Interlinked Datasets} (VoID) \cite{alexander_2009}. VoID is an RDF Schema vocabulary used to describe metadata about RDF datasets such as structural metadata, access metadata and links between datasets. However, VoID is limited regarding terms and design patterns to model the relationships between datasets in a less verbose, unambiguous and explicit way. To overcome this problem, we introduce the {\it concept of virtual link set} (VLS). A virtual link is an intersection data point between two RDF datasets. A data point is any node or resource in an RDF graph such as literals and IRIs (Internationalized Resource Identifier). An RDF dataset is a set of RDF triples that are published, maintained or aggregated by a single provider \cite{alexander_2009}. The links are required in order to comprehend how to semantically relate datasets. The major advantage of the VLS-concept is to facilitate the writing of federated SPARQL queries \cite{harris_2013}, by acting as joint points between the federated sources. We {\it exemplify and apply VoIDext to various world-wide used data sets} and discuss both the theoretical and practical implications of these new concepts with the goal of more easily querying heterogeneous and independent datasets.

This article is structured as follows: Section 2 presents the relevant related work. Section 3 details our approach to extend the VoID vocabulary. In Section 4, we describe the major benefits of using VoIDext, and we apply VoIDext to describe VLSs among three world-wide used bioinformatics RDF data stores. Finally, we conclude this article with future work and perspectives.

\vspace{-10pt}
\section{Related Work}
\vspace{-5pt}
\label{sec_01}

Since the release of the SPARQL 1.1 Query Language \cite{harris_2013} with federated query support in 2013, numerous federated approaches for data and semantic interoperability have recently been proposed \cite{hasnain2017biofed}, \cite{djokic2017pibas}, \cite{Zivanovic_2019}, and \cite{Wimalaratne2015-vi}. However, to the best of our knowledge, none of them proposes a vocabulary and patterns to extensively, explicitly and formally describe how the data sources can be interlinked for the purpose of facilitating the writing of SPARQL 1.1 federated queries such as discussed in Section \ref{sec_02}. In effect, existing approaches put the burden on the SPARQL users or systems to find out precisely \textit{how} to write a conjunctive federated query. An emerging research direction entails automatically discovering links between datasets using Word Embeddings \cite{fernandez2018seeping}. However, the current focus is mostly on relational data or unstructured data \cite{brunner2019entity}. In addition, several link discovery frameworks such as in \cite{Sherif2017-kv}, \cite{Ngomo2011-ea}, and \cite{Isele_undated-ck} rely on link specifications to define the conditions necessary for linking resources within datasets. With these specifications, these frameworks describe similarity measures or distance metrics (e.g. Levenshtein, Jaccard and Cosine) as part of conditions to determine, for example, whether two entities should be linked. The approaches of \cite{fernandez2018seeping}, \cite{Sherif2017-kv}, \cite{Ngomo2011-ea}, and \cite{Isele_undated-ck} are complementary to ours because they can aid in the process of defining virtual link sets---see Def. \ref{def:vlset}. 

In the context of ontology alignment, the Expressive and Declarative Ontology Alignment Language (EDOAL) enables us to represent correspondences between heterogeneous ontological entities \cite{david2011alignment}. Although, transformations of property values can be specified with EDOAL, the current version of EDOAL solely supports a limited kind of transformations\footnote{\scriptsize\url{http://alignapi.gforge.inria.fr/edoal.html}}. In \cite{crotti2017evaluation},  \cite{10.1007/978-3-319-58451-5_3}, and \cite{de2016ontology}, authors also recognize the limited support for data transformation in mapping languages. Moreover, since EDOAL does not focus on supporting the write of SPARQL 1.1 federated queries, the EDOAL data transformation specification requires an extra step to be converted into an equivalent one by using the SPARQL language.  Applying data transformations during a federated query execution is often required to be able to link real-world independent and distributed datasets on the Web. As other related work in terms of RDF-based vocabularies, we can also mention VoID and SPARQL 1.1 Service Description (SD)\footnote{\scriptsize\url{https://www.w3.org/TR/sparql11-service-description/}}. Although the VoID RDF schema provides the \textit{void:Linkset} term (Def. \ref{def:linkset}), this concept alone is not sufficient to precisely and explicitly define virtual links between the datasets (discussed in Section \ref{sec_02}). By precisely, we mean to avoid multiple ways to represent (i.e. triple patterns) and to interpret interlinks. Moreover, by considering Def. \ref{def:linkset} extracted from the VoID specification, this definition impedes the use of the \textit{void:Linkset} concept to describe a link set between instances of the same class because both are triple subjects stored in different datasets. 

\begin{definition}[link set -- \textit{void:Linkset}\footnote{\scriptsize\url{http://vocab.deri.ie/void\#Linkset}}] 
\label{def:linkset}
A collection of RDF links between two datasets. An RDF link is an RDF triple whose subject and object are described in different datasets \cite{alexander_2009}. 
\end{definition}

\vspace{-10pt}
\section{Contribution}
\vspace{-5pt}
\label{sec_02}
To mitigate the impediments of interoperating with distributed and independent RDF datasets, we first propose design patterns of how to partially model virtual links (see Def. \ref{def:vlset}) with the current VoID vocabulary and expose its drawbacks. To address these drawbacks, we then propose a new vocabulary (i.e. VoIDext) and demonstrate an unambiguous and unique way to extensively and explicitly describe various types of virtual links such as depicted in Subsection \ref{subsec:cl_voidext}. The VoIDext vocabulary is fully described in \cite{voidext_io}, that also includes examples of virtual link types.    
\begin{definition}[virtual link set]
\label{def:vlset}
A set of virtual links. A virtual link is a connection
between common resources such as literals and instances from two different RDF datasets. Semantic relaxation (see Def. \ref{def:semantiRelaxation}) is also considered when identifying common resources between datasets.  
\end{definition}
\vspace{-7pt}
\begin{definition}[semantic relaxation]
\label{def:semantiRelaxation}
It is the capacity of ignoring semantic and data heterogeneities for the sake of interoperability.   
\end{definition}

In this article, the words {\it vocabulary} and {\it ontology} are used interchangeably. The methodology applied to develop VoIDext was inspired by the simplified agile methodology for ontology development (SAMOD) \cite{peroni_2017}. Indeed, the proposed VoID extension is a meta-ontology to explicitly describe interlinks between RDF datasets --- virtual links. Further information about how VoIDext was built is given in the Supplementary Material in \cite{voidext_git}.

Fig. \ref{VL_Fig1} illustrates a complex virtual link about Swiss cantons between the LINDAS dataset (Linked Data Service\footnote{\label{lindas}\scriptsize\url{https://lindas-data.ch}}) of the Swiss Government administration and DBpedia \cite{lehmann2015dbpedia}. To define this link, some semantic relaxation is applied. This is because heterogeneities are exacerbated when interoperating independent datasets. For example, what is considered a long name of a Swiss canton in LINDAS is actually a short name in DBpedia. In addition, the data types for the name of the canton are not the same in both datasets what impedes exact matching when performing a federated join query. Finally, LINDAS contains a few literals with different concatenated translations of the same canton's name such as ``Graub\"unden / Grigioni / Grischun'' that can be matched with the literal ``Grisons'' asserted as a canton's short name in DBpedia. Indeed, Grison is the French translation of Graub\"unden --- German name. Nevertheless, both datasets share literals with some commonality. By exploring this commonality we are able to define a virtual link set between both datasets. Note that the Swiss cantons' resource IRIs in both datasets are not the same -- otherwise defining a virtual link set would be simpler --- i.e. a simple link set, see Def. \ref{def_simplelinkset}. In the next subsections, we incrementally demonstrate with a running example how to model a complex link set (see Def. \ref{def_complexlinkset}) with VoID and VoIDext terms. Tab. \ref{tab_01} shows other datasets and SPARQL endpoints considered in our examples in this article.

\begin{table}[h]\scriptsize \centering

\caption{SPARQL endpoints considered in this article.}\label{tab_01} 
\begin{tabular}{ |l|l| }
\hline
	\textbf{RDF Dataset} & \textbf{SPARQL endpoint} \\ \hline
	DBpedia \cite{lehmann2015dbpedia} & http://dbpedia.org/sparql  \\ \hline
	LINDAS\textsuperscript{\ref{lindas}} & https://lindas-data.ch/sparql  \\ \hline
	OMA \cite{oma_2018} & https://sparql.omabrowser.org/sparql  \\ \hline
	UniProtKB\cite{uniProt_Consortium_2018}  & https://sparql.uniprot.org/sparql  \\ \hline
	Bgee\cite{bgee_2019} & http://biosoda.expasy.org/rdf4j-server/repositories/bgeelight \\ \hline
	EBI RDF\cite{ebi_2014} & https://www.ebi.ac.uk/rdf/services/sparql  \\ \hline
\end{tabular}
\end{table}

\begin{figure}[h]
\centerline{\includegraphics[width=1.05\columnwidth]{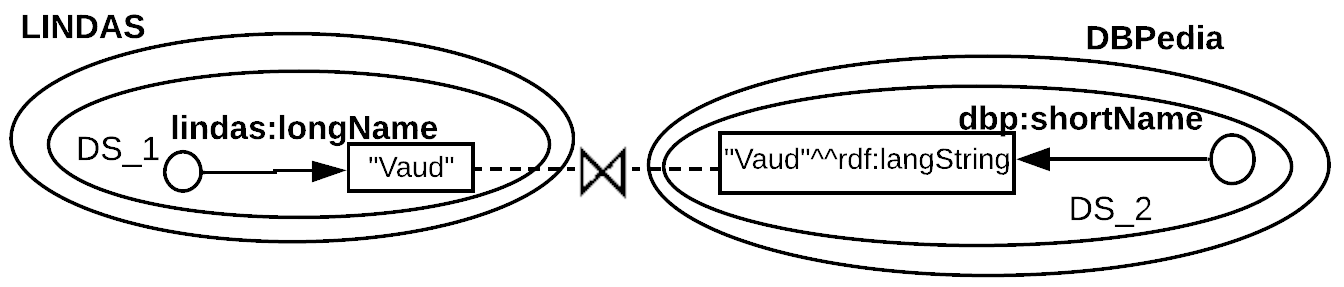}}
\caption{An example of a virtual link between the LINDAS and DBpedia datasets where the $DS_1$ and $DS_2$ datasets are subsets of them, respectively. Circles: different resource IRIs; rectangles: literals; $\bowtie$: virtual link; and edges: RDF predicates.}\label{VL_Fig1}
\end{figure}
\begin{definition}[simple link set]
\label{def_simplelinkset}
A simple link set must be either a link set that does not target another link set (i.e. it has exactly one link predicate --- Def. \ref{def_linkpredicate}) or a set with exactly the same shared instances of the same type (i.e. class expression) in both datasets. 
\end{definition}

\begin{figure}[h]
\centerline{\includegraphics[width=0.85\columnwidth]{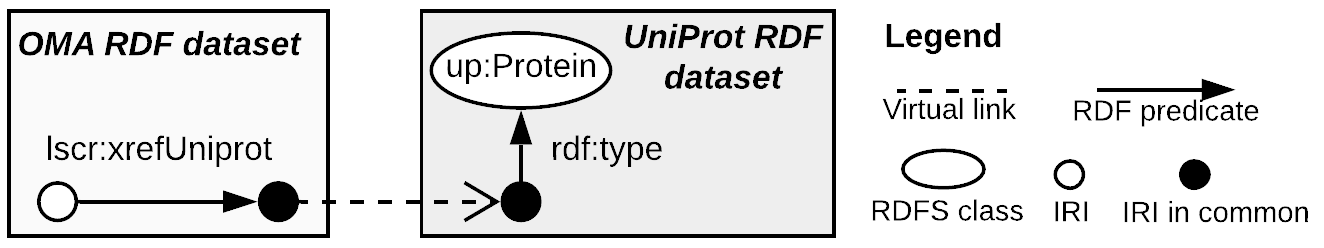}}
\vspace{-10pt}\caption{A simple virtual link between the OMA and UniProt datasets where \textit{lscr:xrefUniprot} is an example of a link predicate.}\label{fig_01}
\end{figure}

In practice, a simple link set allows us to model virtual links either between the subjects of two RDF triples in different datasets where their predicate is \textit{rdf:type} with the same object or between link predicate assertions and  \textit{rdf:type} triples as illustrated in Fig. \ref{fig_01}. Due to the space constraints, the patterns to model simple link sets with VoIDext are available in Supplementary Material Section 4 \cite{voidext_git}.

\begin{definition}[link predicate]
\label{def_linkpredicate}
According to the VoID specification, a link predicate is the RDF property of the triples in a \textit{void:Linkset}\cite{alexander_2009}.   
\end{definition}
\vspace{-5pt}

\vspace{-5pt}

\begin{definition}[complex link set]
\label{def_complexlinkset}
It is a complex virtual link set. A complex link set is composed of exactly two link sets xor two shared instance sets (see Def. \ref{def_sharedinstanceset}) where xor is the exclusive or.
\end{definition}
\vspace{-5pt}
\begin{definition}[shared instance set]
\label{def_sharedinstanceset}
A shared instance set between exactly two datasets. For example, two datasets that contain the same OWL/RDFS class instances.
\end{definition}

\vspace{-15pt}
\subsection{Patterns to \textit{partially} model complex link sets with VoID terms}
\label{subsec_31}
Since our main goal is to facilitate the writing of federated queries by providing metadata of how the target datasets can be joined, let us suppose that we want to know how to relate Swiss cantons in LINDAS and DBpedia datasets as shown in Fig. \ref{VL_Fig1}. In other words, we want to find out the necessary and sufficient graph pattern in the context of Swiss cantons in each dataset to be able to relate and join them. Further triple patterns such as  attributes (e.g. canton's population, cities, acronym) depend on the specificity of the requested information what goes beyond the task of joining the two datasets. Let us further assume a SPARQL user without any previous knowledge about these datasets. A possible workflow for this user to find out how to relate LINDAS and DBpedia in terms of Swiss cantons is described as follows:

\begin{enumerate}[label=\textbf{\arabic*}),topsep=0.3em,leftmargin=0.5em,itemindent=1.2em,labelwidth=\itemindent,labelsep=0em,align=left]
    \item the user has to dig up the data schema and documentation, if any, looking for the abstract entity ``Swiss canton''. This task has to be done for both datasets. 
    \item if (s)he is lucky, a concept is explicitly defined in the data schema. This is the case of the LINDAS dataset that contains the class \textit{lindas:Canton} --- prefixes such as \textit{lindas:} are defined in Tab. \ref{tab_02}. Otherwise the user has to initiate a fastidious quest for assertions and terms that can be used for modeling Swiss canton data. This is the situation of DBpedia where instances are defined as a Swiss canton by assigning the \textit{dbrc:Cantons\_of\_Switzerland} instance of the \textit{skos:Concept} to the \textit{dct:subject} property such as the following triple (\textit{dbr:Vaud, dct:subject, dbrc:Cantons\_of\_Switzerland}).  
    \item  The user has now to browse the RDF graph. For example, by performing additional queries, to be sure that the assertions to the \textit{lindas:Canton} instances can be used as join points with assertions related to Swiss canton instances in DBpedia. Otherwise, the user has to repeat the previous steps.
    \item If data transformations are required because of data and semantic heterogeneities between the datasets, the user has to define data mappings to be able to effectively perform a federated conjunctive query.  
    \end{enumerate}

\begin{table}[ht]
 \vspace{-4mm}
\tiny \centering 
\caption{In this article, we assume the namespace prefix bindings in this table.}\label{tab_02} 
\begin{tabular}{ |l|l| }
\hline
	\textbf{Prefix} & \textbf{Namespace Internationalized Resource Identifier (IRI)} \\ \hline
	rdfs: & http://www.w3.org/2000/01/rdf-schema\#  \\ \hline
	rdf: & http://www.w3.org/1999/02/22-rdf-syntax-ns\#  \\ \hline
	orth: & http://purl.org/net/orth\#  \\ \hline
	up: & http://purl.uniprot.org/core/ \\ \hline
	oboowl: & http://www.geneontology.org/formats/oboInOwl\# \\ \hline
	cco: & http://rdf.ebi.ac.uk/terms/chembl\#  \\ \hline
	chembl: & http://rdf.ebi.ac.uk/resource/chembl/molecule/  \\ \hline
	ex: & http://example.org/voidext\#  \\ \hline
	dbo: & http://dbpedia.org/ontology/  \\ \hline
	skos: & http://www.w3.org/2004/02/skos/core\#  \\ \hline
	dbr: & http://dbpedia.org/resource/ \\ \hline
	dbrc: & http://dbpedia.org/resource/Category: \\ \hline
	dbp: & http://dbpedia.org/property/ \\ \hline
	lindas: & https://gont.ch/ \\ \hline
	dcterms: & http://purl.org/dc/terms/  \\ \hline
    biopax: & http://www.biopax.org/release/biopax-level3.owl\# \\ \hline
	lscr: & http://purl.org/lscr\#  \\ \hline
	void: &  http://rdfs.org/ns/void\# \\ \hline
	voidext: & http://purl.org/query/voidext\#  \\ \hline
	bioquery: & http://purl.org/query/bioquery\#  \\ \hline
\end{tabular}
\end{table}

Finally, based on that workflow, a \textbf{SPARQL user} can draft the minimum set of triple patterns and data transformations to perform the virtual links concerning Swiss cantons between both datasets. This draft is represented as the SPARQL query in Listing \ref{lst_00}. The link set built by intersecting the resources (i.e. the values of \textit{lidas:longName} and \textit{dbp:shortName} properties) can then be partially modelled with VoID terms. This enables \textbf{other SPARQL users or systems} to reuse this link set knowledge to write specialized queries over the two datasets starting from the Swiss canton context. In doing so, the second user avoid the fastidious, complex and time-consuming task of finding this link set. In addition, to the best of our knowledge there is no system capable of precisely establishing this virtual link set automatically because of the complexity and heterogeneities to be solved.   

\begin{lstlisting}[label={lst_00},inputencoding=utf8/latin1,language=SPARQL2 , linewidth=\columnwidth,breaklines=true,captionpos=b,caption=The initial basic graph patterns represented as a SPARQL federated query to perform join queries between Swiss cantons in the LINDAS and DBpedia datasets. Tab. \ref{tab_02} contains the IRI prefixes.,frame=single, basicstyle=\scriptsize\sffamily,float,framextopmargin=0pt,framexbottommargin=0pt,float] 
SELECT * WHERE { 
SERVICE <http://dbpedia.org/sparql>{
?dbp_inst dct:subject dbrc:Cantons_of_Switzerland.
?dbp_inst dbp:shortName ?dbp_name.
BIND(IF(STR(?dbp_name)="Grisons", "Graubünden / Grigioni / Grischun",
    IF(STR(?dbp_name)="Geneva", "Genève",
    IF(STR(?dbp_name)="Lucerne", "Luzern",
    IF(STR(?dbp_name)="Valais", "Valais / Wallis",
    IF(STR(?dbp_name)="Bern", "Bern / Berne",
    IF(STR(?dbp_name)="Fribourg", "Fribourg / Freiburg", 
        STR(?dbp_name) )))))) AS ?lindas_name)}
SERVICE <https://lindas-data.ch/sparql>{
?lindas_inst a  lindas:Canton;
             lindas:longName ?lindas_name.}}
\end{lstlisting}

Listings \ref{VL_Fig2} and \ref{VL_Fig3} depict two different ways named $VL_{m1}$ and $VL_{m2}$ to model the virtual link set with VoID. Note that the examples of RDF graph patterns in this section are defined with the RDF 1.1 Turtle language\footnote{\scriptsize\url{https://www.w3.org/TR/turtle/}}. On the one hand, $VL_{m1}$ states that a given $LS_1$ link set targets another $LS_2$ link set that targets $LS_1$ back. On the other,  $VL_{m2}$ only states datasets as link set targets. By using the $VL_{m1}$ model in Listing \ref{VL_Fig2}, the $DS_2$ instance asserts the $DS_1$ instance to its \textit{void:objectsTarget} property. As a reminder, the \textit{void:objectsTarget} value is the dataset describing the objects of the triples contained in the link set, in our example, the objects of \textit{dbp:shortName}. This dataset must contain only the relevant triples to describe the virtual link set. In our example in Listing \ref{VL_Fig2}, we define the $DS_1$ dataset (i.e. a subset of LINDAS) as being also a \textit{void:Linkset} that contains triples with the  \textit{lindas:longName} predicate. By using the $VL_{m2}$ model in Listing \ref{VL_Fig3}, the objects' target dataset of the \textit{dbp:shortName} link predicate is not a \textit{void:Linkset} but a \textit{void:Dataset} (i.e. superclass of \textit{void:Linkset}). $DS_1$ in Listing \ref{VL_Fig3} also contains triples with the  \textit{lindas:longName} predicate, however, this predicate is defined as part of a subset and partition of $DS_1$ by using the \textit{void:propertyPartition} and \textit{void:property} terms. Note that solely one \textit{void:propertyPartition} should be directly assigned to $DS_1$ dataset, otherwise we are not able to know which predicate should be considered when stating the virtual links.

\begin{table}[!ht]
\renewcommand\tablename{Listing} 
\noindent\begin{minipage}{.49\textwidth}
\begin{lstlisting}[frame=tb, inputencoding=utf8/latin1,language=SPARQL2 , linewidth=\columnwidth,breaklines=true, basicstyle=\scriptsize\sffamily,escapechar=\%,framextopmargin=0pt,framexbottommargin=0pt]{Name}
%\textbf{\#DS\_1 is a subset of LINDAS.}%
%ex:LINDAS\_DBPEDIA\_SWISSCANTON% 
  %\textbf{rdf:type} void:Linkset%;
  %\underline{\textbf{void:linkPredicate} lindas:longName}%;
  %\textbf{void:objectsTarget} ex:DBPEDIA\_LINDAS\_SWISSCANTON%;
  %\textbf{void:subset} \_:DOMAIN\_SET0%.
%\_:DOMAIN\_SET0 \textbf{void:propertyPartition} \_:b0%;
  %\textbf{void:classPartition} \_:b1 .%
%\_:b0 \textbf{void:property} \textit{rdfs:domain} .%
%\_:b1 \textbf{void:class} \textit{lindas:Canton} .%
                    %\dots%
\end{lstlisting}
\end{minipage}\hfill 
\begin{minipage}{.49\textwidth}
\begin{lstlisting}[frame=tb, inputencoding=utf8/latin1,language=SPARQL2 , linewidth=\columnwidth,breaklines=true, basicstyle=\scriptsize\sffamily,escapechar=\%,framextopmargin=0pt,framexbottommargin=0pt]{Name}
%\textbf{\#DS\_2 is a subset of DBpedia.}%
%ex:DBPEDIA\_LINDAS\_SWISSCANTON% 
  %\textbf{rdf:type} void:Linkset%;
  %\underline{\textbf{void:linkPredicate} dbp:shortName}%;
  %\textbf{void:objectsTarget} ex:LINDAS\_DBPEDIA\_SWISSCANTON%;
  %\textbf{void:subset} \_:DOMAIN\_SET1 .%
%\_:RANGE\_SET1 \textbf{void:propertyPartition} \_:b2%;
  %\textbf{void:classPartition} \_:b3 .%
%\_:b2 \textbf{void:property} \textit{rdfs:range} .%
%\_:b3 \textbf{void:class} \textit{rdf:langString} .%
                    %\dots%
\end{lstlisting}
\end{minipage}
  \caption{Patterns to model a complex virtual link set between the LINDAS Linked Data service and DBpedia relying on link sets as targets (e.g. \textit{void:objectsTarget}). As a reminder, \textit{void:Linkset} is a subclass of \textit{void:Dataset}. }
    \label{VL_Fig2}
     \vspace{-4mm}
\end{table}

\begin{table}[!ht]
\centering
\renewcommand\tablename{Listing} 
\begin{minipage}{0.8\textwidth}
\begin{lstlisting}[frame=tb, inputencoding=utf8/latin1,language=SPARQL2 , linewidth=0.8\columnwidth,breaklines=true, basicstyle=\scriptsize\sffamily,escapechar=\%,framextopmargin=0pt,framexbottommargin=0pt]{Name}
%\textbf{\#DS\_2 is a subset of DBpedia.}%
%ex:DBPEDIA\_LINDAS\_SWISSCANTON \textbf{rdf:type} void:Linkset%;
  %\underline{\textbf{void:linkPredicate} dbp:shortName}%;
  %\textbf{void:objectsTarget} \_:b4%.
%\_:b4 \textbf{rdf:type} void:Dataset%;
  %\textbf{void:propertyPartition} \_:LINDAS\_PROPERTY%;
  %\textbf{void:subset} \_:RANGE\_SET2%.
%\_:LINDAS\_PROPERTY \textbf{void:property}  \emph{lindas:longName}%.
%\_:RANGE\_SET2 \textbf{void:propertyPartition}  \_:b5.%
%\_:RANGE\_SET2 \textbf{void:classPartition} \_:b6 .%
  %\_:b5 \textbf{void:property} \emph{rdfs:range}%.    %\#it restricts lindas:longName% 
  %\_:b6 \textbf{void:class} \emph{rdfs:Literal}%.      %\#range to rdfs:Literal.%
                     %\dots%
\end{lstlisting}
\end{minipage}
  \caption{Patterns to model a complex virtual link set between the LINDAS Linked Data service and DBpedia  relying on link sets and property partitions as targets (e.g. \textit{void:objectsTarget}). For the sake of simplicity, only the link set in DBpedia is depicted because the link set in LINDAS containing the \textit{lindas:longName} link predicate is similarly modelled as the one in DBpedia.}
    \label{VL_Fig3}
     \vspace{-4mm}
\end{table}

Yet, we also need to describe further information about the virtual link set such as the domain and range of the link predicates (e.g. \textit{lindas:longName} and \textit{dbp:shortName}). This information is used to restrict which resource type must be considered for a given triple that contains the link predicate (e.g. \textit{lindas:longName rdfs:domain lindas:Canton}). By having this information in advance when writing and executing a federated query, we reduce the number of triples to match, if there are statements of the same predicate but with resources of other types. For example, the \textit{lindas:longName} property is asserted to instances of \textit{lindas:MunicipalityVersion}, \textit{lindas:Canton} or \textit{lindas:DistrictEntityVersion}. However, for the context of this virtual link set only \textit{lindas:Canton} instances need to be considered. To restrict the resource types of a given link predicate with VoID, we can state subsets and partitions to a \textit{void:Linkset} such as exemplified in Listing \ref{VL_Fig2} by using $VL_{m1}$. 

Note that for each link predicate's domain/range, we have to create one new subset to be sure that the class partitions of the subset correspond to the  domain or range of the link predicate. In addition, if there are multiple resource types to be considered as the domain of a link predicate, we can state multiple class partitions to express the union of types --- i.e. classes. Or, we can explicitly define it by using the OWL 2 Description Logic (DL) term \textit{owl:unionOf} and related patterns to express class union. To express class intersection or other class expressions, we can rely on OWL 2 DL terms and state these class expressions as class partitions of the subset. Similarly, we can model the domain and range of predicates related to virtual links with  $VL_{m2}$ as described in Listing \ref{VL_Fig3}. For the sake of simplicity, we do not depict all predicate domains/ranges in Listings \ref{VL_Fig2} and \ref{VL_Fig3}.     

However, there are several limitations when only considering VoID terms to model complex link sets. 

\begin{enumerate}[label=\textbf{\arabic*}),topsep=0.3em,leftmargin=0.5em,itemindent=1.2em,labelwidth=\itemindent,labelsep=0em,align=left]
    \item \textbf{Multiple representations}. The VoID vocabulary and documentation due to the lack of constraints and high generalization imply several ways to model virtual link sets such as $VL_{m1}$ and $VL_{m2}$ graph patterns to represent a complex link set. In addition, there are various ways to define the link predicate's domain and range. For example, class expressions can be defined by using either OWL 2 DL terms to express the union of classes or multiple class partitions (i.e. \textit{void:classPartition} assertions), or by combining both of them. This multitude of graph patterns allowed by VoID makes interoperability more complex because we do not previously know how the virtual link set is modelled. Consequently, it requires to build complex parsers and queries to retrieve the virtual link set metadata.    
    \item \textbf{Ambiguity}. With VoID, we cannot easily distinguish if a link set or dataset is being instantiated to define a virtual link set. For example, we do not know explicitly if two link sets compose a complex link set. Moreover, the use of property/class partitions to define domains and ranges of link predicates can be mixed with \textit{void:class} assertions that are not part of a domain/range definition. Subsets can also be arbitrarily stated to any link set or dataset what increases the ambiguity to know if a given subset is actually part of a complex link set definition or not. With $VL_{m1}$ and $VL_{m2}$ models strictly based on VoID, we cannot explicitly state that the intersections between two link sets occur by matching the subjects-objects, objects-objects or subjects-subjects of link predicates in different link sets. Nevertheless, this information can be derived from the \textit{void:objectsTarget} and \textit{void:subjectsTarget} assertions, if any. 
    \item \textbf{Description Logic (DL) compliance} \cite{baader2003description}. By stating the domain and range of link predicates with class expressions based on OWL 2 DL (e.g. a range composed of multiple types/classes), we can take advantage of existing DL-parser and reasoner tools\footnote{\scriptsize\url{http://owlcs.github.io/owlapi/}} to infer instance types. However, since we can mix DL-based class expressions with \textit{void:class} assertions, the resulting range/domain expressions are non-compliant with DL.
    \item \textbf{Verbosity}. The use of class and property partition partners considerably increases the number of triples to state for representing virtual links. This also increases the complexity of writing of queries to retrieve the virtual link set metadata.
    \item \textbf{Resource mapping}. VoID does not provide any explicit term and recommendation to state resource mappings. By doing so, we mitigate or even solve heterogeneities when matching resources with some commonality in different datasets. 
\end{enumerate}

In the next subsection, we show how to solve these issues with VoIDext terms and patterns.

\vspace{-10pt}
\subsection{Patterns to \textit{fully} model complex link sets with VoIDext}
\label{subsec:cl_voidext}
To address the issues of modelling virtual link sets solely with VoID, we propose new terms and patterns in VoIDext. Listing \ref{VL_Fig4} illustrates the main  VoIDext terms (see terms with \textit{voidext:} prefix) and design patterns to model complex link sets. To assert the range and domain of predicates with VoIDext, we can directly assign the \textit{voidext:linkPredicateRange}  and \textit{voidext:linkPredicateDomain} properties to a link set, respectively (see Def. \ref{def_linkpredicaterange} and Def. \ref{def_linkpredicatedom}). Complex link predicates' domains and ranges (e.g. multiple types --- union/intersection of classes) must be stated as class expressions by using OWL 2 DL terms (e.g. \textit{owl:unionOf}). To avoid ambiguities when interpreting link sets (i.e. a simple set \textit{versus} a complex one), we can explicitly state that two link sets are indeed part of a complex link set. To do so, we must assign exactly two link sets to a complex link set with the \textit{voidext:intersectAt} property (see Def. \ref{def_intersectat}). In a complex link set, a link set must be connected to another link set by stating either \textit{void:objectsTarget} or \textit{void:subjectsTarget} properties. This allow us to precisely know where the intersection between RDF triples with predicates in different datasets occurs, in other words, the matched RDF resource nodes: object-object, subject-subject, and subject-object. For example, in Listing \ref{VL_Fig4} with \textit{void:objectsTarget} property, we state that the \textit{lindas:longName} predicate's objects in LINDAS match the objects of the \textit{dbp:shortName} link predicate in DBpedia, and \textit{vice-versa}. To explicitly state the intersection type (e.g. object-object), we can assert the \textit{voidext:intersectionType} property (see Def. \ref{def_intersectiontype}) to a complex link set as shown in Listing \ref{VL_Fig4}. 

\vspace{-5pt}

\begin{definition}[link predicate range]
\label{def_linkpredicaterange}
The link predicate's object type (i.e. class expression or literals), if any. Moreover, a link set (Def. \ref{def:linkset}) that is not part of a complex link set (see Def. \ref{def_complexlinkset}) and connects two datasets through the link predicate's object must specify the link predicate range. Indeed, this object matches a second resource in another dataset. Therefore, the type of this second resource is asserted as the link predicate range.  
\end{definition}

\begin{definition}[link predicate domain]
\label{def_linkpredicatedom}
The link predicate's subject type (i.e. class expression), if any.  
\end{definition}

\begin{definition}[intersects at]
\label{def_intersectat}
It specifies the intersection of either exactly two shared instance sets (see Def. \ref{def_sharedinstanceset}) or two link sets, that compose a complex link set. 
\end{definition}

\begin{definition}[intersection type]
\label{def_intersectiontype}
It specifies the intersection type between two RDF triples in different datasets. In other words, if the intersection occurs at the subject xor the object node of a link predicate.
\end{definition}

Based on Def. \ref{def_complexlinkset}, the \textit{voidext:ComplexLinkSet} OWL class is defined with the following DL expression, IRI prefixes are ignored to improve readability: 
\begin{align*}
ComplexLinkSet \equiv \neg SimpleLinkSet\: \sqcap \\(( 
\forall intersectAt.Linkset\: \sqcap\:  =2\: intersectAt )\: \sqcup \\(\forall  intersectAt.SharedInstanceSet\: \sqcap\:  =2\: intersectAt))
\end{align*}

As a reminder, a \textit{void:Dataset} is ``a set of RDF triples that is published, maintained or aggregated by a single provider"\footnote{\scriptsize \url{http://vocab.deri.ie/void\#Dataset}}. However, a complex link set is composed of resources from two different datasets (e.g. two link predicates). Therefore, we define \textit{voidext:ComplexLinkSet} as being disjoint with \textit{void:Dataset} class. Consequently, a complex link set is not a \textit{void:Dataset} and properties such as \textit{void:propertyPartition} cannot be assigned to it.

\begin{table}[!ht]
\renewcommand\tablename{Listing} 
\noindent\begin{minipage}{.39\textwidth}
\begin{lstlisting}[frame=tb, inputencoding=utf8/latin1,language=SPARQL2 , linewidth=\columnwidth,breaklines=true, basicstyle=\scriptsize\sffamily,escapechar=\%,framextopmargin=0pt,framexbottommargin=0pt]{Name}
%\textbf{\#DS\_1 is a subset of LINDAS.}%
%ex:LINDAS\_DBPEDIA\_SWISSCANTON%
  %\textbf{rdf:type} void:Linkset%;
  %\underline{\textbf{void:linkPredicate} lindas:longName}%;
  %\textbf{voidext:linkPredicateRange} rdfs:Literal%;
  %\textbf{voidext:linkPredicateDomain} lindas:Canton%;
  %\textbf{voidext:isSubsetOf} ex:LINDAS;%
  %\textbf{void:objectsTarget}%
 %ex:DBPEDIA\_LINDAS\_SWISSCANTON%;
  %\dashuline{\textbf{voidext:resourceMapping}}%
    %```?x a  $<$https://gont.ch/Canton$>$. \dots '{}''%;
            %\dots%
\end{lstlisting}
\end{minipage}\hfill 
\begin{minipage}{.595\textwidth}
\begin{lstlisting}[frame=tb, inputencoding=utf8/latin1,language=SPARQL2 , linewidth=\columnwidth,breaklines=true, basicstyle=\scriptsize\sffamily,escapechar=\%,framextopmargin=0pt,framexbottommargin=0pt, numbers=right, stepnumber=1,  firstnumber=1, numberfirstline=true]{Name}
%\textbf{\#DS\_2 is a subset of DBpedia.}%
%ex:DBPEDIA\_LINDAS\_SWISSCANTON%
  %\textbf{rdf:type} void:Linkset%;
  %\underline{\textbf{void:linkPredicate} dbp:shortName}%;
  %\textbf{voidext:isSubsetOf} ex:DBPEDIA%;
  %\textbf{void:objectsTarget} ex:LINDAS\_DBPEDIA\_SWISSCANTON%; 
  %\dashuline{\textbf{voidext:resourceMapping}}% 
  %```?dbpedia\_place dcterms:subject dbrc:Cantons\_of\_Switzerland.%
  %?dbpedia\_place dbp:shortName ?c.%
  %BIND(%
   %IF(STR(?c)=``Grisons", ``Graub\"unden / Grigioni / Grischun",%
   %IF(STR(?c)=``Geneva", ``Gen\`eve",%
   %IF(STR(?c)=``Lucerne", ``Luzern",%
   %IF(STR(?c)=``Valais", ``Valais / Wallis",%
   %IF(STR(?c)=``Bern", ``Bern / Berne",%
   %IF(STR(?c)=``Fribourg", ``Fribourg / Freiburg", STR(?c) %
      %)))))) as ?lindas\_objects)'{}'';      \dots%
\end{lstlisting}
\end{minipage}
\vspace{-8pt}
\begin{lstlisting}[frame=tb, inputencoding=utf8/latin1,language=SPARQL2 , linewidth=\columnwidth,breaklines=true, basicstyle=\scriptsize\sffamily,escapechar=\%,framextopmargin=0pt,framexbottommargin=0pt]{Name}
%ex:DBPEDIA\_LINDAS\_SWISSCANTON\_VL \underline{\textbf{rdf:type} voidext:ComplexLinkSet}%;
  %\textbf{voidext:intersectionType} voidext:OBJECT\_OBJECT%;
  %\textbf{dcterms:issued}  ``2019-06-30"\textasciicircum\textasciicircum xsd:date%;
  %\textbf{rdfs:label} ``A virtual link set for cantons in both DBpedia and LINDAS Swiss government datasets."%;
  %\textbf{voidext:intersectAt} ex:LINDAS\_DBPEDIA\_SWISSCANTON%;
  %\textbf{voidext:intersectAt} ex:DBPEDIA\_LINDAS\_SWISSCANTON%;
  %\textbf{voidext:recommendedMapping} ex:LINDAS\_DBPEDIA\_SWISSCANTON%;    %\dots%
\end{lstlisting}

\caption{VoIDext-based patterns to model a complex virtual link set between the LINDAS Linked Data service and DBpedia relying on link sets as targets. Dashed underlined: one of the two can be chosen as the \textit{voidext:recommendedMapping}; and fully underlined: predicates used to connect the datasets by \textit{void:objectsTarget} predicates. }
    \label{VL_Fig4}
     \vspace{-4mm}
\end{table}

To address data heterogeneities, we can implement semantic relaxation by stating the \textit{voidext:resourceMapping} property (Def. \ref{def_resourcemapping}) with a literal text based on SPARQL language. In Listing \ref{VL_Fig4}, $DS_2$ states a mapping in line 7 that converts \textit{dbp:shortName} language-tagged string values into simple literals and maps the values to a corresponding one in LINDAS dataset. Thus, since this mapping is defined using SPARQL language, it can be directly used to build a SPARQL 1.1 federated query to perform the interlinks between datasets. In Listing \ref{VL_Fig4}, \textit{voidext:recommendedMapping} (Def. \ref{def_rresourcemapping}) assigns the LINDAS $DS_1$ link set as the one containing the mapping to be considered when interlinking with DBpedia in the context of Swiss cantons. 

\begin{definition}[resource mapping]
\label{def_resourcemapping}
It specifies the mapping function ($f_m$) to preprocess a resource (i.e. IRI or literal) in a source dataset in order to match another resource in the target dataset. The resource preprocessing (i.e. mapping) must be defined with the SPARQL language by mainly using SPARQL built-ins for assignments (e.g. BIND), and expression and testing values (e.g. IF and FILTER). The BIND built-in is used to assign the output of $f_m$, if any. 
\end{definition}

\begin{definition}[recommended resource mapping]
\label{def_rresourcemapping}
It specifies one recommended mapping function, if more than one mapping is defined in the different sets that are part of a complex link set. 
\end{definition}

To exemplify a complex link set composed of shared instance sets (Def. \ref{def_sharedinstanceset}), let us consider the UniProt and European Bioinformatics Institute (EBI) RDF datasets (see Tab. \ref{tab_01}). EBI and UniProt RDF data stores use different instance IRIs and classes to represent the organism species, and in a more general way, the taxonomic lineage for organisms. To exemplify this, let us consider the \textit{\seqsplit{\textless http://identifiers.org/taxonomy/9606\textgreater}} instance of \textit{biopax:BioSource} and the \textit{\seqsplit{\textless http://purl.uniprot.org/taxonomy/9606\textgreater}} instance of \textit{up:Taxon} in EBI and UniProt datasets, respectively. Although these instances are not exactly the same (i.e. distinct IRIs, property sets, and contexts), they refer to the same organism species at some extent, namely \textit{homo sapiens} --- human. By applying a semantic relaxation, we can state a virtual link between these two instances. To establish this link, we need to define a resource mapping function (i.e. $f_m(r)$) either to the EBI or UniProt species-related instances ---  either  $f_m(<$\textit{http://identifiers.org/taxonomy/9606}$>) \equiv <$\textit{\seqsplit{http://purl.uniprot.org/taxonomy/9606}}$>$ or $f_m(<$\textit{\seqsplit{http://purl.uniprot.org/taxonomy/9606}}$>) \equiv <$\textit{\seqsplit{http://identifiers.org/taxonomy/9606}}$>$. Listing \ref{VL_Fig5} depicts how this complex link set is modelled with VoIDext-based patterns. Note that it is not possible to define a shared instance set by only using VoID terms because there is no link predicate (Def. \ref{def_linkpredicate}) associated with the interlinks that are different from \textit{rdf:type}. To address this issue, we can assign a shared instance type (Def. \ref{def_sharedinstancetype}) with the \textit{voidext:sharedInstanceType}  property for each \textit{voidext:SharedInstanceSet} instance (Def. \ref{def_sharedinstanceset}). Other examples of complex link sets are available in \cite{voidext_git} and \cite{voidext_io}.

\begin{definition}[shared instance type]
\label{def_sharedinstancetype}
The type (i.e. class) of the shared instances in a given dataset. Shared instances imply equivalent or similar instance IRIs that belong to different datasets.
\end{definition}

\vspace{-12pt}

\section{VoIDext benefits and discussions}
 VoID instances (``assertion box'' --- ABox) are fully backward compatible with the VoIDext schema since we mainly add new terms. The only change performed  in the VoID ``terminological box'' (TBox) concerns the \textit{void:target}\footnote{\scriptsize\url{https://www.w3.org/TR/void/\#target}\label{fn_target}} property domain. In VoIDext, this domain is the union of the \textit{void:Linkset} and \textit{voidext:SharedInstanceSet} classes instead of solely \textit{void:Linkset}, as stated in VoID. We did this to avoid the replication of a similar property to state target datasets to shared instance sets. Despite this modification, assertions of \textit{void:target} based on VoID remain compatible with VoIDext.   

\begin{table}[h]
\renewcommand\tablename{Listing} 
\noindent\begin{minipage}{.5\textwidth}
\begin{lstlisting}[frame=tb, inputencoding=utf8/latin1,language=SPARQL2 , linewidth=\columnwidth,breaklines=true, basicstyle=\scriptsize\sffamily,escapechar=\%,framextopmargin=0pt,framexbottommargin=0pt]{Name}
%\textbf{A) a subset of EBI}%
%bioquery:EBI\_UNIPROT\_10 \underline{\textbf{rdf:type voidext:SharedInstanceSet}}%;
  %\textbf{voidext:isSubsetOf} bioquery:EBI%;
  %\dashuline{\textbf{voidext:resourceMapping}}% 
  %```?IRI\_EBI a biopax:BioSource.%
%BIND(IRI(CONCAT(``http://purl.uniprot.org/taxonomy/"%
  %, STRAFTER(STR(?IRI\_EBI),%
  %``http://identifiers.org/taxonomy/"))) as ?IRI\_UNIPROT)%
  %FILTER(STRSTARTS(STR(?IRI\_EBI),%
  %``http://identifiers.org/taxonomy/"))'{}'';%
  %\textbf{voidext:sharedInstanceType} biopax:BioSource%; %\dots%
\end{lstlisting}
\end{minipage}\hfill 
\begin{minipage}{.5\textwidth}
\begin{lstlisting}[frame=tb, inputencoding=utf8/latin1,language=SPARQL2 , linewidth=\columnwidth,breaklines=true, basicstyle=\scriptsize\sffamily,escapechar=\%,framextopmargin=0pt,framexbottommargin=0pt]{Name}
%\textbf{B) a subset of UniProt}%
  %bioquery:EBI\_UNIPROT\_11 \underline{\textbf{rdf:type voidext:SharedInstanceSet}}%;
  %\textbf{voidext:isSubsetOf} bioquery:UNIPROT%;
  %\dashuline{\textbf{voidext:resourceMapping}}%
  %```?IRI\_UNIPROT a up:Taxon.%
  %BIND(IRI(CONCAT(``http://identifiers.org/taxonomy/"%
  %, STRAFTER( STR(?IRI\_UNIPROT),%
  %``http://purl.uniprot.org/taxonomy/") ) ) as ?IRI\_EBI)%
  %FILTER(STRSTARTS(STR(?IRI\_UNIPROT),%
  %``http://purl.uniprot.org/taxonomy/"))'{}''%;
  %\textbf{voidext:sharedInstanceType} up:Taxon%; %\dots%
\end{lstlisting}
\end{minipage}
\vspace{-12pt}
\begin{lstlisting}[frame=tb, inputencoding=utf8/latin1,language=SPARQL2 , linewidth=\columnwidth,breaklines=true, basicstyle=\scriptsize\sffamily,escapechar=\%,framextopmargin=0pt,framexbottommargin=0pt]{Name}
%bioquery:EBI\_UNIPROT\_12 \textbf{rdf:type voidext:ComplexLinkSet}%;
  %\textbf{voidext:intersectAt} bioquery:EBI\_UNIPROT\_11%;
  %\textbf{voidext:intersectAt} bioquery:EBI\_UNIPROT\_10%;
  %\textbf{voidext:recommendedMapping} bioquery:EBI\_UNIPROT\_10%;
%\textbf{rdfs:label} ``Links between EBI and UniProt considering shared similar instances of organism taxonomy"@en;% %\dots%  
\end{lstlisting}
\caption{VoIDext-based patterns to model a complex virtual link set between EBI and UniProt datasets modelled with shared instance sets (see fully underlined assertions). Dashed underlined: one of the two can be chosen as the \textit{voidext:recommendedMapping}. }
    \label{VL_Fig5}
    \vspace{-4mm}
\end{table}

\subsection{Retrieving virtual link sets}
\label{queryset}

Once the virtual links are modelled with VoIDext as discussed in Subsection \ref{subsec:cl_voidext} and Supplementary Material Section 4 \cite{voidext_git}, there may exist at most four kinds of virtual link sets as follows: (i) a \textit{voidext:ComplexLinkSet} composed of \textit{void:Linkset}s --- e.g. see Listing \ref{VL_Fig4}; (ii) a \textit{voidext:ComplexLinkSet} composed of \textit{voidext:SharedInstanceSet}s --- e.g. see Listing \ref{VL_Fig5}; (iii) a \textit{void:Linkset} that is also a \textit{voidext:SimpleLinkSet}; and (iv) a \textit{voidext:SharedInstanceSet} that is also a \textit{voidext:SimpleLinkSet}. Due to the page limit, the types (iii) and (iv) are exemplified in Fig. 7 and Listing 1.3 in Supplementary Material \cite{voidext_git}. For each kind of virtual link set, a SPARQL query template to retrieve the essential information is asserted as an annotation of the \textit{voidext:ComplexLinkSet} and \textit{voidext:Simple\-Link\-Set} sub-classes of \textit{voidext:VirtualLinkSet}. These annotations are done  by asserting the   \textit{voidext:query\-Linkset} and \textit{voidext:querySharedInstanceSet} properties. Therefore, to retrieve virtual link sets of type (i) and (iii), we can execute the SPARQL queries assigned with \textit{voidext:queryLinkset} to the \textit{voidext:Complex\-Link\-Set} and \textit{voidext:Simple\-Lin\-k\-Set} classes, respectively. Similarly, to retrieve virtual link sets of type (ii) and (iv), we can execute the SPARQL queries assigned with \textit{voidext:query\-Shared\-Instance\-Set} to the \textit{voidext:Complex\-Link\-Set} and \textit{voidext:Simple\-Link\-Set} classes, respectively.  These queries are described in \cite{voidext_io}. 

\vspace{-7pt}
\subsection{Writing a federated SPARQL query with VoIDext metadata }
To illustrate how VoIDext can facilitate the writing of federated SPARQL queries, let us consider that a SPARQL user wants to perform the $Q_f$ query against the EBI dataset: \textit{``Show me all assays in rodents for the drug Gleevec (i.e. CHEMBL941 identifier)"}. One possible way to write this query is to consider another dataset that contains organismal taxonomy information about rodents such as the UniProt dataset. To be able to write this federated conjunctive query over the EBI and UniProt datasets, the SPARQL user has to find out how to relate these datasets. To do so, the SPARQL user can query the metadata about virtual link sets modelled with VoIDext---see the template queries for this purpose in the VoIDext specification \cite{voidext_io} and further details in Subsection \ref{queryset}. Tab. \ref{tab_05} exemplifies a possible outcome of these queries containing virtual link set descriptions. 

Based on the description of virtual link sets, users can select the link set that best fit their needs to write $Q_f$ with SPARQL 1.1. In this example, a user can choose the complex link set \textit{bioquery:EBI\_UNIPROT\_12} about organism taxonomy depicted in Listing \ref{VL_Fig5} and defined as the query result over the VoIDext metadata in Tab. \ref{tab_05}---i.e. the $T_1$ tuple. By considering $T_1$ in Tab. \ref{tab_05}, the SPARQL user can draft $Q_f$ starting with the interlink between EBI and UniProt as shown in bold in Listing \ref{qf}. The user can now continue the writing of $Q_f$ by solely focusing on each dataset separately---i.e. the non-bold part of the $Q_f$ SPARQL query. Therefore, the fastidious process of finding out interlinks and data transformations between EBI and UniProt to perform a federated query is mitigated with the virtual link sets defined using the VoIDext vocabulary. The query in Listing \ref{qf} can be executed in the EBI SPARQL endpoint (see Tab. \ref{tab_01}). Further examples of SPARQL federated queries that were written based on VoIDext metadata are available as part of an application described in Subsection \ref{subsec_43}.

\begingroup
\setlength{\tabcolsep}{1pt} 
\begin{table}[h!]\scriptsize \centering
\caption{The results of querying complex link sets composed of two shared instance sets between EBI and UniProt. }\label{tab_05}
\setlength{\extrarowheight}{2pt}{
\begin{tabular}{ |p{12.2cm}| }
\hline
	\textbf{Outcome:}  A set of tuples $T = (V_{L}, {ds}_1, {ds}_2,  I_{t1}, I_{t2}, A_{ds_1}, A_{ds_2}, f_m)$ where $V_{L}$ is the virtual link set IRI; ${ds}_1$ and ${ds}_2$ are the names of the datasets that contain the instance IRIs; $I_{t1}$ is the type (DL-class expression) of the instance in ${ds}_1$; $I_{t2}$ is the type (DL-class expression) of the instance in ${ds}_2$; $A_{ds_1}$ and $A_{ds_2}$ are the access methods such as SPARQL endpoints to the ${ds}_1$ and ${ds}_2$ datasets, respectively; $f_m$ is the recommended resource mapping procedure, if any, to be applied to instance IRIs of $I_{t1}$ xor $I_{t2}$ types, where xor is the exclusive or.  \\ \hline 
	\textbf{Example:}  \begin{center} $T_1=($bioquery:EBI\_UNIPROT\_12, ``Linked Open Data platform for EBI data.",``The Universal Protein Resource (UniProt)", biopax:BioSource, up:Taxon,  $<$https://www.ebi.ac.uk/rdf/services/sparql$>$,  $<$https://sparql.uniprot.org/sparql/$>$, $f_m^3(i)$)
	\end{center} 
	where $i$ is any instance of \textit{biopax:BioSource} type and $f_m^3(i)\equiv$ \newline  
```\texttt{?IRI\char`_EBI} a $<$http://www.biopax.org/release/biopax-level3.owl\#BioSource$>$.\newline
BIND(IRI(CONCAT(``http://purl.uniprot.org/taxonomy/", STRAFTER( \newline
STR(\texttt{?IRI\char`_EBI}), ``http://identifiers.org/taxonomy/"))) as \texttt{?IRI\char`_UNIPROT})\newline 
FILTER(STRSTARTS(STR(\texttt{?IRI\char`_EBI}), ``http://identifiers.org/taxonomy/"))'''   \\ \hline
\end{tabular}
}
\end{table} 
\endgroup

\begin{lstlisting}[label={qf},inputencoding=utf8/latin1,language=SPARQL2 , linewidth=\columnwidth,breaklines=true,captionpos=b,caption=A federated query between EBI and UniProt datasets to retrieve assays in rodents for the drug Gleevec (i.e. CHEMBL941). Tab. \ref{tab_02} contains the IRI prefixes.,frame=single, basicstyle=\scriptsize\sffamily,escapechar=\%,float,framextopmargin=0pt,framexbottommargin=0pt] 
SELECT ?assay WHERE { 
 ?activity a cco:Activity ;
          cco:hasMolecule chembl:CHEMBL941 ;
          cco:hasAssay ?assay .
 ?assay cco:taxonomy  ?IRI_EBI.

 %\textbf{?IRI\_EBI a $<$http://www.biopax.org/release/biopax-level3.owl\#BioSource$>$ .}%
 %\textbf{BIND(IRI(CONCAT(``http://purl.uniprot.org/taxonomy/",}%
 %         \textbf{STRAFTER(STR(?IRI\_EBI),``http://identifiers.org/taxonomy/"))) as ?IRI\_UNIPROT)}%
 %\textbf{FILTER(STRSTARTS(STR(?IRI\_EBI),``http://identifiers.org/taxonomy/"))}%
         
 %\textbf{SERVICE$<$https://sparql.uniprot.org/sparql$>$\{}%
  %\textbf{?IRI\_UNIPROT a up:Taxon. }%
  ?IRI_UNIPROT rdfs:subClassOf ?taxon2  .
  ?taxon2 up:otherName %``%rodents". } }
\end{lstlisting}

\vspace{-10pt}
\subsection{Virtual link set maintenance}
Although, to manage the virtual link set evolution  is out of the scope of this article, we recommend to annotate the link sets with the issued and modified dates such as depicted in Listing \ref{VL_Fig4}. This date information helps with the maintenance of virtual link sets. For example, let us suppose the release of a new version of the DBpedia in August, 2019. By checking the difference between the DBpedia new release date and the complex link set issued/modified date (e.g. June 2019, see Listing \ref{VL_Fig4}), it might indicate a possible decrease in the virtual link set performance, or even, invalidity of the interlinks due to the fact of being outdated. In addition, for each virtual link set, we can state the performance in terms of precision, recall, true positives, and so on by asserting the \textit{voidext:hasPerformanceMeasure} property. The range of this property is \textit{mex-perf:PerformanceMeasure}\textsuperscript{\ref{mex}}. Thus, we can rely on the Mex-perf ontology \footnote{\scriptsize\label{mex}\url{http://mex.aksw.org/mex-perf}} to describe the virtual link set performances. The complex link set example about Swiss cantons in Listing \ref{VL_Fig4} has a precision and recall of 100\%. In this example, for every Swiss canton in LINDAS exists a corresponding one in DBpedia. Therefore, if this performance is deteriorated after the new release of one of the datasets involved, we should review this virtual link set. Further use cases are exemplified in Supplementary Material Section 5 available in \cite{voidext_git}.

\vspace{-10pt}
\subsection{Benefits and a SIB Swiss Institute of Bioinformatics' application}
\label{subsec_43}
\vspace{-3pt}
\paragraph{\textbf{Easing the task of writing SPARQL 1.1 federated queries.}}
The formal description of virtual link sets among multiple RDF datasets on the Web facilitates the manually or (semi-)automatically writing of federated queries. This is because once the virtual link sets are defined between datasets with VoIDext, we can interlink different RDF datasets without requiring to mine this information again from the various ABoxes and TBoxes (including documentation, if any). The mining task becomes more and more complex and fastidious if the TBox is incomplete or missing when comparing with the ABox statements, for example, a triple predicate that is not defined in the TBox.  

\vspace{-5pt}
\paragraph{\textbf{Applying semantic relaxation rather than semantic reconciliation.}}
The \textit{virtual link} statements between datasets are more focused on the meaning of interlinking RDF graph nodes rather than the semantics of each node in the different datasets and knowledge domains. For example, let us consider the \textit{virtual link} illustrated in Fig. \ref{VL_Fig1}. When considering solely the LINDAS dataset, the \textit{lindas:longName} is a \textit{rdf:Property} labelled as a ``District name or official municipality name". In DBpedia, \textit{dbp:shortName} is a \textit{rdf:Property} labelled as ``short name'' and in principle can be applied to any instance. Hence, it is not restricted to district names. In addition, one property is about long names while the other one is about short names. However, they state similar literals in the context of Swiss cantons as discussed in Section \ref{sec_02}. Therefore, although these properties are semantically different (hard to reconcile), we can still ignore heterogeneities for the sake of  interlinking DBpedia and LINDAS.  

\vspace{-6pt}
\paragraph{\textbf{Facilitating knowledge discovery.}}
As noticed in \cite{acosta_2018}, yet there are many challenges to address in the semantic web such as the previous knowledge of the existing RDF datasets and how to combine them to process a query. VoIDext mitigates these issues because RDF publishers (including third-party ones) are able to provide virtual link sets which explicitly describe how heterogeneous datasets of distinct domains are related. Without knowing these links, to potentially extract new knowledge that combines these datasets is harder or not even possible. The virtual link sets stated with VoIDext terms provide sufficient machine-readable information to relate the datasets. Nonetheless, the automatic generation of these link sets is out of the scope of this article.
\vspace{-6pt}
\paragraph{\textbf{A SIB Swiss Institute of Bioinformatics application.}} We applied the VoIDext vocabulary in the context of a real case application mainly involving three in production life-sciences datasets available on the Web, namely UniProtKB, OMA and Bgee RDF stores --- see SPARQL endpoints in Tab. \ref{tab_01}. The RDF serialization of virtual link sets among these three databases is available in \cite{voidext_git} and it can be queried via the SPARQL endpoint in \cite{sparql_endpoint_vl} with query templates defined in \cite{voidext_io} as described in Subsection \ref{queryset}. Based on these \textit{virtual links}, a set of more than twelve specialized federated query templates over these data stores was defined and are available at  {\scriptsize\url{https://github.com/biosoda/bioquery/tree/master/Queries}}. These templates are also available through a template-based search engine, see {\scriptsize\url{http://biosoda.expasy.org}}. Moreover, as an example of facilitating knowledge discovery, we can mention the virtual link sets between OMA and Bgee. These two distinct biological knowledge domains when combined enable to predict gene expression conservation for orthologous genes (i.e. corresponding genes in different species). Finally, new virtual link sets are being created to support other biological databases in the context of SIB --- {\scriptsize\url{https://sib.swiss}}.
\vspace{-8pt}
\section{Conclusion}
\vspace{-6pt}
We successfully extended the VoID vocabulary (i.e. VoIDext) to be able to formally describe \textit{virtual links} and we provided a set of SPARQL query templates to retrieve them. To do so, we applied an agile methodology based on the SAMOD approach. We described the benefits of defining \textit{virtual links} with VoIDext RDF schema, notably to facilitate the writing of federated queries and knowledge discovery. In addition, with  \textit{virtual links} we can enable interoperability among different knowledge domains without imposing any changes in the original RDF datasets. In the future, we intend to use VoIDext to enhance keyword-search engines over multiple distributed and independent RDF datasets. We also envisage to propose tools to semi-automatically  create VoIDext virtual link statements between RDF datasets. We believe these tools can leverage the adoption of VoIDext by other communities besides SIB, Quest for Orthologs consortium ({\scriptsize\url{https://questfororthologs.org}}), and Linked Building Data Community ({\scriptsize\url{https://www.w3.org/community/lbd}}) where the authors are involved. We also encourage other communities to collaborate on open issues in the public GitHub of VoIDext in \cite{voidext_git} to refine this vocabulary for other use cases that have not been contemplated during this work. Finally, to support virtual link evolution, we  aim to develop a tool to automatically detect broken virtual links because of either data schema changes or radical modifications of instances' IRIs and property assertions.

\vspace{5pt}
\noindent{\textbf{Acknowledgements.}}
This work was funded by the Swiss National Research Programme 75 ``Big Data'' (Grant 167149) and a Swiss National Science Foundation Professorship grant to CD (Grant 150654).
\vspace{-10pt}
\bibliographystyle{splncs04}
\bibliography{ref.bib}
\end{document}